# Understanding the fate of corona virus transmission using a simple model

Proma Mondal*, Timothy A. Gonsalves and Aniruddha Chakraborty
School of Basic Sciences, Indian Institute of Technology Mandi, Himachal Pradesh 175075, India.

We propose a simple model for understanding the kinetics of corona virus transmission. Our model assume spreading of corona virus can happen from one to another only, if someone without enough protection comes close contact to a person carrying the corona virus. Therefore this virus spreads on a large scale within a short time through chains of such events. Using our model we provide an estimation of the number of people affected by this virus within reasonable duration of time. We choose values of different parameters of our model by non-linear least square fit of the real time data and we predict fate of this corona virus transmission using our model.

The current corona-virus outbreak is an ongoing global outbreak of corona-virus disease. The outbreak was first identified in Wuhan, China in 31-st of December 2019, and has since been declared as Public Health Emergency of international concern by the World Health Organization [1, 2]. As of 9-th March 2020, over 114000 cases have been confirmed in the entire world [4]. More than 110 countries, with major outbreaks in central China, Italy, South Korea and Iran [3]. More than 4000 people have died, among those more than 3100 in China and around 900 in other countries [3]. As of now no vaccine or specific anti-viral treatment is available, though research is going on. Efforts are aimed at managing symptoms and supportive therapy. Recommended preventive measures include maintaining distance from other people (particularly those who are sick) [7, 8] as it is understood that this virus spreads between people via respiratory droplets from coughing or sneezing [6]. Therefore research on corona virus has become an urgent topic with practical significance. In this paper we look at the theoretical aspect of this problem. As very less information is available on corona virus even today, in the following we assume spreading of corona virus can happen from one to another only, if someone without enough protection comes close contact to a person carrying the corona virus. Therefore this virus spreads on a large scale within a short time through chains of such events. Using this model we estimate the spread and possible cessation of corona virus transmission. As the detailed mechanism of corona-virus transmission is not known yet, therefore this is the only model possible. In our model, people are divided into three classes: ignorant (those not aware of this virus -there number should be large at the initial stage) and let x(t) be the number of such people at time t, spreaders (those who have this virus and are spreading it -this number should be small at the initial stage) and let y(t) be the number of such people at time t) and stiflers (those who have this virus but are not spreading it -they are isolated or under medication or dead -this number should be smallest at the initial stage) and let z(t) be the number of such people at time t. All these three sub-classes of people together are called 'corona virus' class and we expect this class to expand in time and we assume all these additional people belongs to 'ignorant' sub-class. There will be another class of individuals who will not be an explicit part of our model are those who will never meet spreader or even if they meet spreader nothing will happen to them (know how to protect themselves or are lucky not to meet any spreader or are genetically protected!). Similar models are already available in literature in the context of rumour transmission [9, 10]. In our model, spreader-ignorant contact will convert the ignorant to spreader -, corona virus will be transmitted from the spreader to the ignorant, spreader-spreader contact will do nothing and the spreader-stifler contact will stifle the spreader (stiflers are placed isolated at the different locations along with doctors, therefore meeting stiflers is equivalent to meeting a doctor ¬hence when a spreader meets doctor, spreader will be isolated and called as stifler). It is important to mention here that increase in the number of stifler is equivalent to increase in number of doctors or hospitals for accommodating more stiflers, therefore spreader gets more chance to become stifler. The total population size at time t is denoted by N(t), where N(t)= x(t)+y(t)+z(t). The model is described by the following system of coupled differential equations

$$\frac{dx}{dt} = \Lambda - \frac{k\,x\,y}{1+\alpha\,y^2} - \mu\,x$$

$$\frac{dy}{dt} = \frac{k\,x\,y}{1+\alpha\,y^2} - \lambda\,y\,z - \mu\,y$$

$$\frac{dz}{dt} = \lambda\,y\,z - \mu\,z \qquad (1)$$

When a spreader contacts an ignorant, the spreader transmits the corona virus with a probability and we assume that g(y) x ignorant change their 'corona-virus' sub-class and becomes spreader during the small time interval from t to t + δt, where g(y) is the incident rate, and is given by $g(y) = ky/(1 + \alpha y^2)$. In general it is quite natural to assume the incident rate (the rate of new infections) to be linear in the infected number y. At low values of y incident rate can be assumed to be linear in y. But at high values of y, the number of 'corona-virus transfer' contacts between the infected individual and ignorant individual may reduce due to protection measures taken by the ignorant individuals. Hence incident rate rate is expected to increase slower than linear in y. Therefore the use of non-monotone function g(y) in the corona-virus transmission model is more realistic. The function g(y) is increasing when y is small and decreasing when y is large. In Eq. (1), ky measures the probability of new corona virus infection and $1/(1 + \alpha y^2)$ measures the inhibition effect from the protection taken by the ignorant individuals. It is important to note that when α = 0, the non-monotonic incident rate becomes the linear incidence rate. The parameter k is the probability of infection due to the corona virus and the parameter α describe the quality of protection taken by the general people against this corona virus infection. Large values of α means more knowledgeable people who can take quality protection against the corona-virus infection. Fig. 1 shows how g(y) changes with y for different values of k (denotes the rate of infection due to corona virus). The peak value of g(y) measures the maximum corona virus influence. The higher, the probability of new infection, the larger the maximum corona-virus influence. The smaller the probability of new infection, corona-virus transmission terminates very quickly. Fig. 2 shows the plot of g(y) changes with y for different values of α (denote the protection taken by susceptible individuals against the corona-virus). For larger values of α, the smaller the maximum is the influence of corona-virus and the influence force decrease to zero within a short period. As we can see that by helping the people with more protection against the corona virus, one can speed up the time at which the virus-transmission reaches it's lowest value. For a smaller value of α, the influence force due to corona virus can have reached their maximum value s subsequently the influence force take very long time to decrease to zero. In Eq. (1) we consider the following, when a spreader contacts a stifler, the spreader is isolated and therefore spreader becomes a stifler. This can be understood by the fact that stiflers are placed in different specialized hospitals and the number of such specialized hospitals increases with increase in number of stifler. With more specialized hospitals i.e., with more expert doctors, the rate at which spreader become stifler increases. Therefore we assume that λyzδt spreader becomes stifler during the small time interval from t and t + δt. In the above Λ denote the number of new people added in the 'corona-virus' class ("births") at all time and μ is the number of people goes out of the 'corona-virus' class ("deaths"). We assume that newcomers in the corona-virus class, all are ignorant and the emigration is independent of the 'corona-virus' sub-class (same rate for all three types), k is the proportionality constant and α is the parameter that measures the 'inhibitory' effect. There are five unknown parameters in our model i.e., Λ, k, α, μ and λ and also the initial conditions i.e., the value of $x(t_0)$, $y(t_0)$ and $z(t_0)$ is not known. Also we do not have any reliable data for number of confirmed corona-virus infected people for first 20 days, reliable data is available from 21-st day onward [5]. Therefore it is really difficult to have a good estimate of the value of $x(t_0)$, $y(t_0)$ and $z(t_0)$. So in the following we use 20-th day as the initial time of our calculation i.e., t = 20 and final time in our simulation is assumed to 500 days, which is reasonably long. Therefore we will integrate the Eq. (1) from t = 20 to t = 500. Now the available real time data [3, 5] reports the number of confirmed cases of corona-virus infected people, which is nothing but number of stifler, because after confirmation of the corona-virus infection, they must have been isolated from the rest of the world -so that they won't be able to spread this corona-virus to anyone, anymore. In the following we assume the initial condition, i.e., the value of x(20), y(20) and z(20) are unknown parameters. Therefore now we have total eight unknown parameters i.e., Λ, k, α, μ, λ, x(20), y(20) and z(20) and we need to estimate the values of all these unknown parameters. This is done by performing a non-linear least square fit of the real time data given in Ref. 5, where we get the daily data of number of corona-virus infected people (with confirmed corona-virus infection). We assume those confirmed people must have been isolated (so that they won't be able to spread corona-virus) -we call them as stifler i.e., z(t) in our model. Therefore, we fit the real time data of Ref. 5 with z(t) of our model. Best fit is possible for the following values of the unknown parameters. i.e. Λ = 31547 per day, k = 156 per day, α = 51.6, μ =5.86 × $10^{-9}$ per day and λ = 2138 per day, with the initial condition x(20) = 88355, y(20) = 8004 and z(20) = 300. These values are used for obtaining results of all our numerical calculations related to spreading of corona-virus transmission

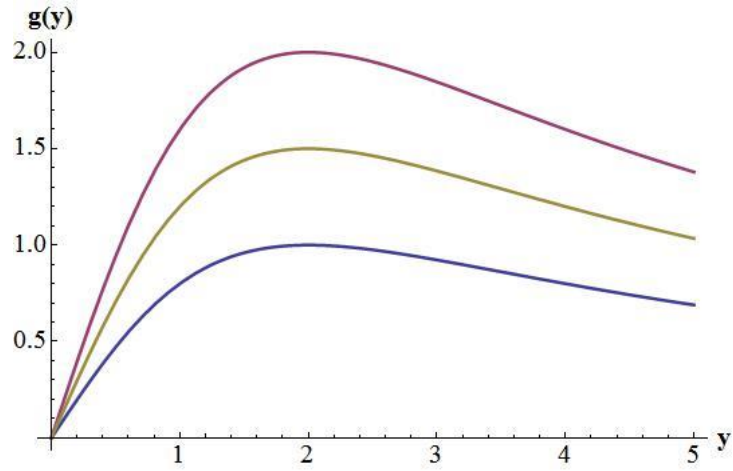

FIG. 1: Plot of incident function g(y) vs. y. This plot is made to show the effect of k values on g(y). All three plots have α =0.25, and the top plot is for k =1.0, middle plot is for k =1.5 and the bottom plot is for k =2.0.

Using Eq. (1) we get an the following equation for the total population N to satisfy

$$\frac{dN}{dt} = \Lambda - \mu N \qquad (2)$$

The solution of the above equation is given by

$$N(t) = N_0 \, e^{-\mu t} \, \frac{1 - \Lambda \, e^{-\mu t}}{\mu} \qquad (3)$$

Now if we assume $N_0 = x_0 + y_0 + z_0 = \Lambda/\mu$, then we will get $N_t = x_t + y_t + z_t = \Lambda/\mu$, that means we will have population of constant size -which is not the case for corona-virus transmission problem. Therefore we will use the initial condition, which does not satisfy the above equation. Now we consider the existence of equilibrium in system defined by Eq. (1). The condition for equilibrium is (x =0, y =0, z = 0) and this leads to the following three equations

$$\Lambda - \frac{k\,x\,y}{1 + \alpha\,y^2} - \mu\,x = 0$$

$$\frac{k\,x\,y}{1 + \alpha\,y^2} - \lambda y\,z - \mu\,y = 0$$

$$\lambda y\,z - \mu\,z = 0$$

(4)

One of the possible solutions of Eq. (4) is the trivial solution given by

$$\{x \to \frac{\Lambda}{\mu}, y \to 0, z \to 0\} \qquad (5)$$

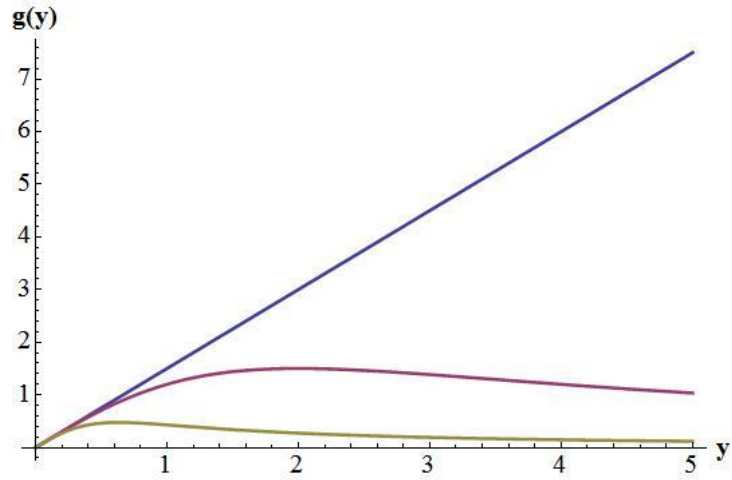

FIG. 2: Plot of incident function g(y) vs. y. This plot is made to show the effect of α values on g(y). All three plots have k =1.5, and the top plot is for α = 0, middle plot is for α =0.25 and the bottom plot is for α =2.5.

-which is the case for corona-virus free equilibrium. This is the case where corona virus was never introduced into the system, therefore this solution is not interesting to us! From this, we can see that there should be no people either spreader or stifler classes. To find the other possible equilibria, we solve the above equation analytically and find the following solutions

$$\{x \to \frac{\frac{\sqrt{-4\alpha\mu^2+k^2+4\alpha k \Lambda}}{\alpha}+\frac{k}{\alpha}+2\Lambda}{2\mu}, y \to \frac{-\mu\sqrt{-4\alpha\mu^2+k^2+4\alpha k \Lambda}-k\mu}{2\alpha\mu^2}, z \to 0\}$$

$$\{x \to \frac{-\frac{\sqrt{-4\alpha\mu^2+k^2+4\alpha k \Lambda}}{\alpha}+\frac{k}{\alpha}+2\Lambda}{2\mu}, y \to \frac{\mu\sqrt{-4\alpha\mu^2+k^2+4\alpha k \Lambda}-k\mu}{2\alpha\mu^2}, z \to 0\} \quad (6)$$

The above two solutions are not accepted physically, since number of stifler is zero i.e., z = 0. Now the other possible solution is given by

$$\{x \to \frac{\alpha\Lambda\mu^2+\lambda^2\Lambda}{\mu(\alpha\mu^2+\lambda^2+k\lambda)}, y \to \frac{\mu}{\lambda}, z \to \frac{\alpha\mu^4-\lambda^2\mu^2+k\lambda^2\Lambda-k\lambda\mu^2}{\lambda\mu(\alpha\mu^2+\lambda^2+k\lambda)}\} \quad (7)$$

This is the only solution where all variables x,y and z are nonzero, therefore it represent corona-virus persistent equilibrium. In our numerical solution we find that all variables reach these equilibrium value after a very long time.

Fig. 3 shows how total number of people in the corona virus class N(t) changes with time. This figure shows that N(t) increases with time and within 200 days approximately 6000000 people will be in the corona-virus class. Fig. 4 shows how total number of ignorant people x(t) in the corona-virus class changes with time. As N(t) increases, x(t) also increases very fast. Almost 5000000 people will be in the ignorant class within 200 days. Fig. 5 shows how total number of spreader people y(t) in corona virus class changes with time. . As N(t) increases, x(t) also increases very fast. Almost 5000000 people will be in the ignorant class within 200 days. Fig. 5 shows how total number of spreader people y(t) in corona virus class changes with time. Within a very short span of time the number of spreader increases very sharply and reaches the maximum possible value and then it decreases and finally it reaches almost zero value within 200 days (with the increase in awareness of the people). One can see from the equilibrium values of y(t) that y(t) approaches zero, as α becomes infinity. The larger value of α the more knowledgeable people, they show good resistance against the spreading of corona-virus infection. The smaller values of α people show fragility to corona-virus transmission. Fig. 6 shows how total number of stifler people z(t) in corona virus class changes with time. Initially this number was very small then it increase because some

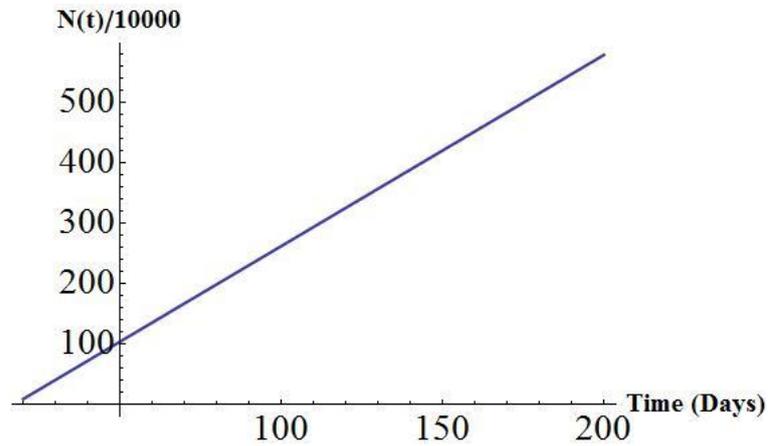

FIG. 3: Plot of total population of corona-virus class N(t) vs. time (in days). The parameter values used in this plot are Λ = 31547 per day, k = 156 per day, α = 51.6, μ = 5.86 × $10^{-9}$ per day and λ = 2138 per day, with the initial condition x(20) = 88355, y(20) = 8004 and z(20) = 300.

- therefore, the incident function approach a small values (less than 5) within 200 days. In Fig. 8, we compare our results with that found in literature [5]. number of stifler is represented by z(t). Hence the total number of corona-virus infected people identified in the entire world should be compared with z(t) of our model and this done in Fig. 8. Fig. 8 shows that our model fits well with the available real time data [5]. Our model is quite general and is applicable to any infectious disease. Different diseases are expected to be modelled by different set of parameter values of Eq.(1). Therefore real time data is required for estimating different parameter values of our model. To validate our model we apply our model for understanding another type of corona-virus transmission i.e., the case of Severe Acute Respiratory Syndrome (SARS). It is a viral respiratory disease caused by the SARS coronavirus (SARS-CoV). Between November 2002 and July 2003, an outbreak of SARS in southern China caused 8, 098 infected cases, resulting in 774 deaths reported in worldwide [13], with most of the cases in China and Hong Kong [14]. No cases of SARS have been reported worldwide since 2004 [14]. We have found real time data for total 95 days starting from 78-th day [12]. We took first 30 real time data and estimated all eight different parameter values of our model and result of our calculation is shown in Fig. 9. Our result shows that rate of new infection approaches zero within 120 days, but in reality it happens after 160 days. But it is important to note that our prediction is based on only 30 data points and it is reasonably good. Now we took first 60 real time data and estimated different parameter values of our model and result of our calculation is shown in Fig. 10. Our result shows that rate of new infection approaches zero within 150 days, but in reality it happens after 160 days. At last we took all 95 real time data and estimated different parameter values of our model and result of our calculation is shown in Fig. 11. Our result shows that rate of new infection approaches zero within 160 days, which is what one expect from real time data. Therefore, we can safely assume that if enough amounts of data points are used for estimating different parameters of our model, then the predictions using our model will be reasonably good. We proposed one simple model to understand the transmission of corona virus. Our model explains real time data very well. Different parameters of our model, is estimated using the real time data taken from reliable resource [5]. Therefore the prediction based on our model is expected to be reliable. Our model predicts that the rate of new corona-virus infection will reduced to less than 5 per day, after approximately 200 days.

Data Availability Statement
Data will be made available on request from the authors.

ACKNOWLEDGMENTS

One of the authors (P.M.) would like to thank IIT Mandi for HTRA and the other author (A.C.) would like to thank IIT Mandi for PDA.

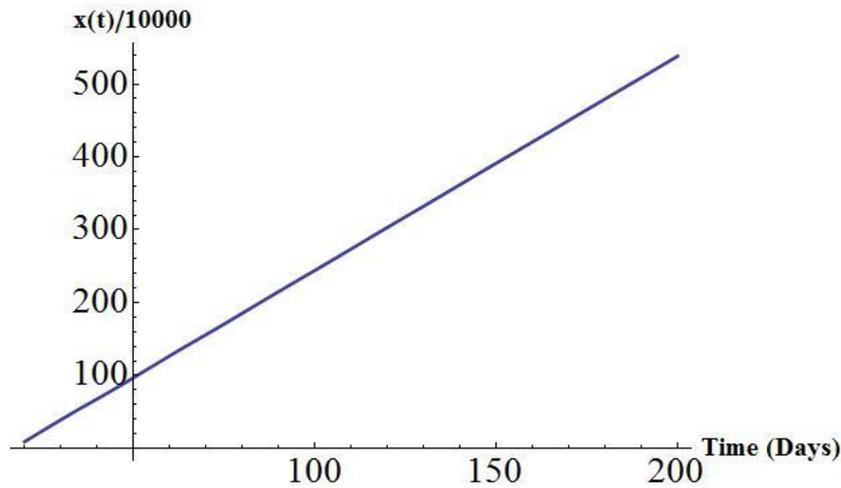

FIG. 4: Plot of number of 'ignorant' vs. time (in days) for corona-virus class. The parameter values used in this plot are Λ = 31547 per day, k = 156 per day, α = 51.6, μ =5.86 × $10^{-9}$ per day and λ = 2138 per day, with the initial condition x(20) = 88355, y(20) = 8004 and z(20) = 300.

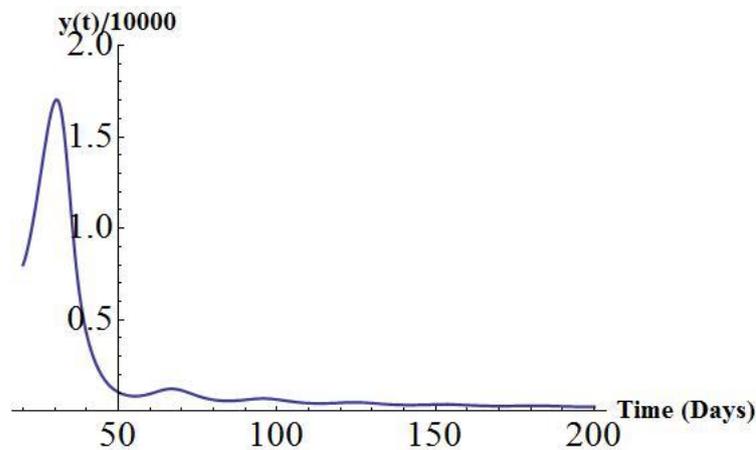

FIG. 5: Plot of number of 'spreader' vs. time (in days) for corona-virus class. The parameter values used in this plot are Λ = 31547 per day, k = 156 per day, α = 51.6, μ =5.86 × $10^{-9}$ per day and λ = 2138 per day, with the initial condition x(20) = 88355, y(20) = 8004 and z(20) = 300.

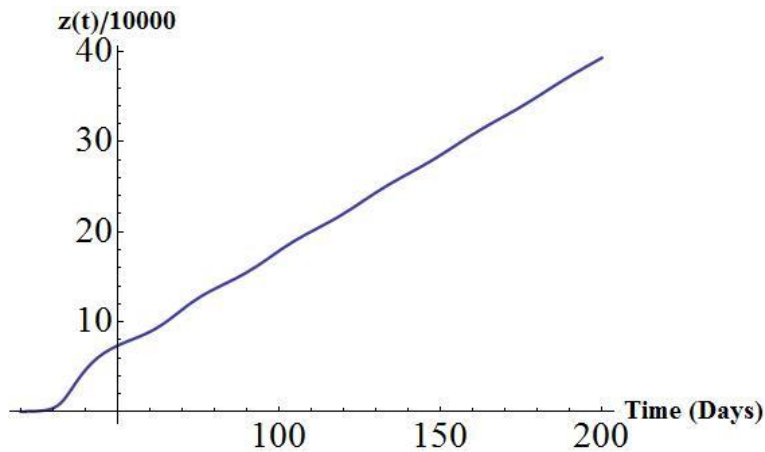

FIG. 6: Plot of number of 'stifler' vs. time (in days) for corona-virus class. The parameter values used in this plot are Λ = 31547 per day, k = 156 per day, α = 51.6, μ = 5.86 × $10^{-9}$ per day and λ = 2138 per day, with the initial condition x(20) = 88355, y(20) = 8004 and z(20) = 300.

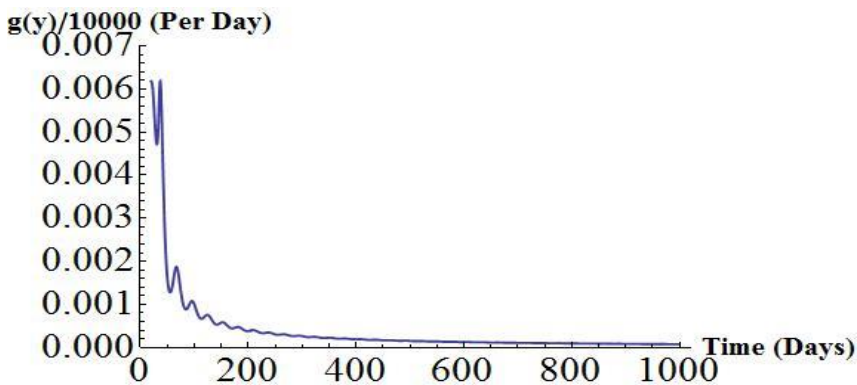

FIG. 7: Plot of incident function g[y(t)] vs. time (in days) for corona virus class. The parameter values used in this plot are Λ = 31547 per day, k = 156 per day, α = 51.6, μ = 5.86 × $10^{-9}$ per day and λ = 2138 per day, with the initial condition x(20) = 88355, y(20) = 8004 and z(20) = 300.

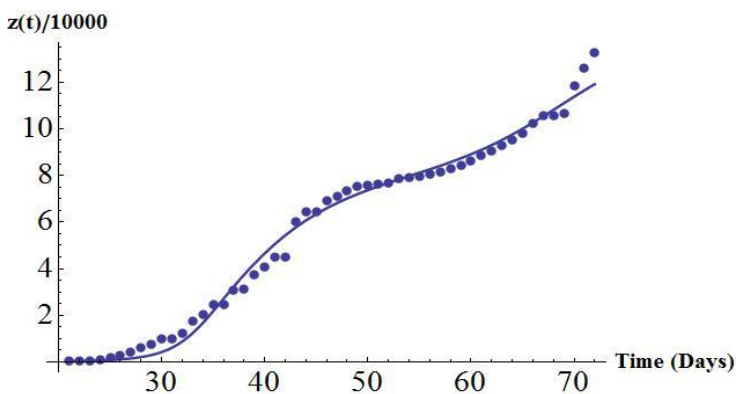

FIG. 8: Plot showing the comparison real data (dots) available in Ref. 4 and the results (z(t) vs. t) using our model (solid line. The parameter values used in this plot are Λ = 31547 per day, k = 156 per day, α = 51.6, μ = 5.86 × $10^{-9}$ per day and λ = 2138 per day, with the initial condition x(20) = 88355, y(20) = 8004 and z(20) = 300.

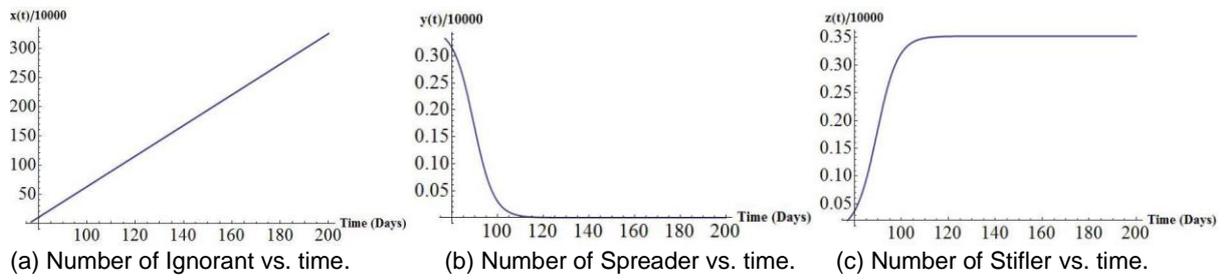

(a) Number of Ignorant vs. time.   (b) Number of Spreader vs. time.   (c) Number of Stifler vs. time.

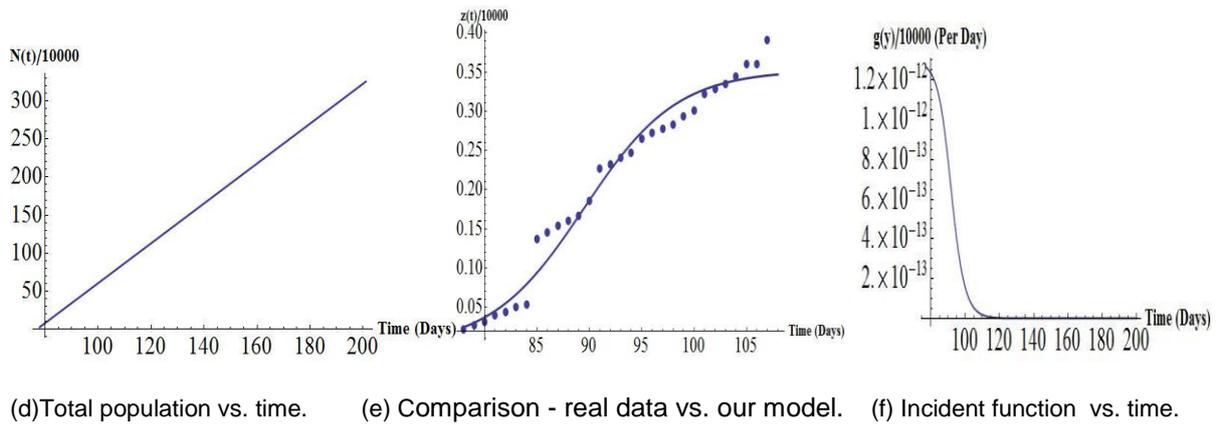

(d) Total population vs. time.   (e) Comparison - real data vs. our model.   (f) Incident function vs. time.

FIG. 9: Results using our model for SARS. The parameter values used in all the plots are $\Lambda = 26243$ per day, $k = 5.31^{-8}$ per day, $\alpha = 3.56$, $\mu = 2.16^{-8}$ per day and $\lambda = 6392$ per day, with the initial condition $x(77) = 22721$, $y(77) = 3322$ and $z(77) = 200$. These parameters are generated by fitting 30 real time data points of Ref. 6.

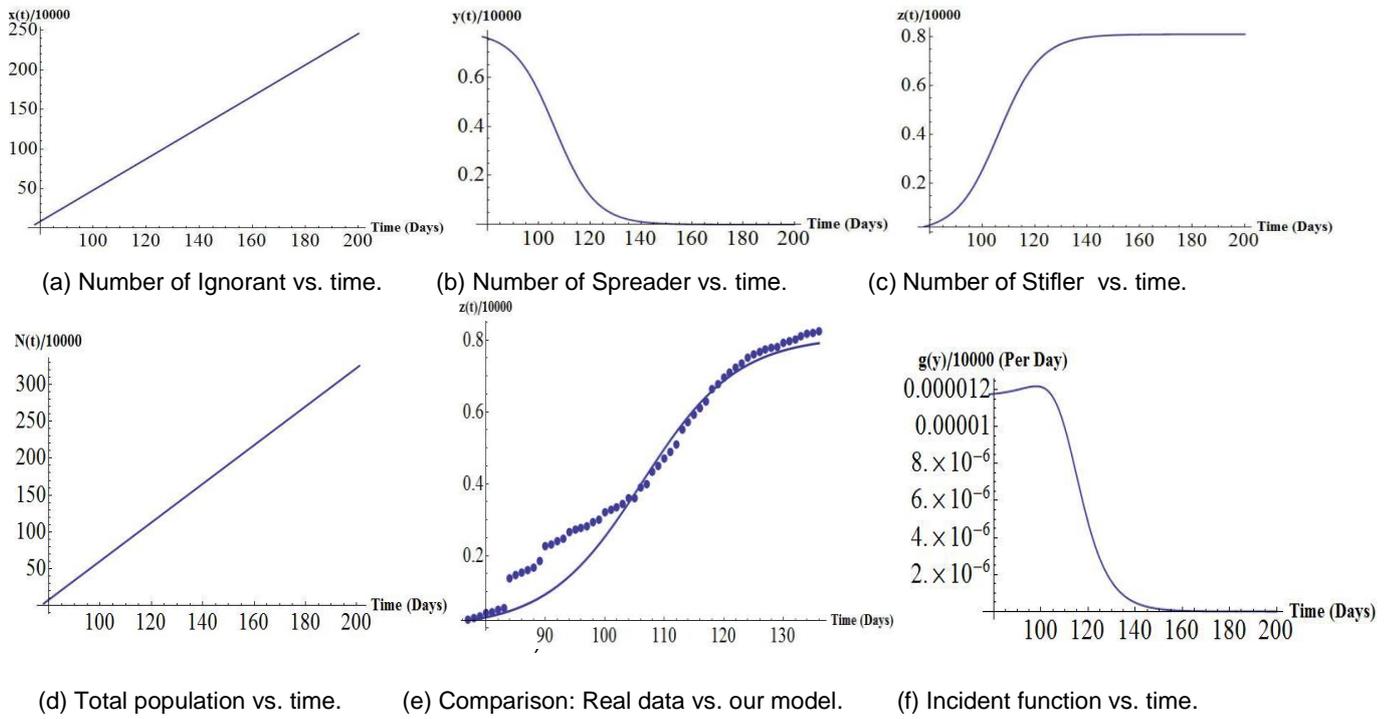

(a) Number of Ignorant vs. time.  (b) Number of Spreader vs. time.  (c) Number of Stifler vs. time.

(d) Total population vs. time.  (e) Comparison: Real data vs. our model.  (f) Incident function vs. time.

FIG. 10: Results using our model for SARS. The parameter values used in all the plots are $\Lambda = 19786$ per day, $k = 0.48$ per day, $\alpha = 2.94$, $\mu = 7.93 \times 10^{-5}$ per day and $\lambda = 1596$ per day, with the initial condition $x(77) = 24192$, $y(77) = 7669$ and $z(77) = 200$. These parameters are generated by fitting 60 real time data points of Ref. 6.

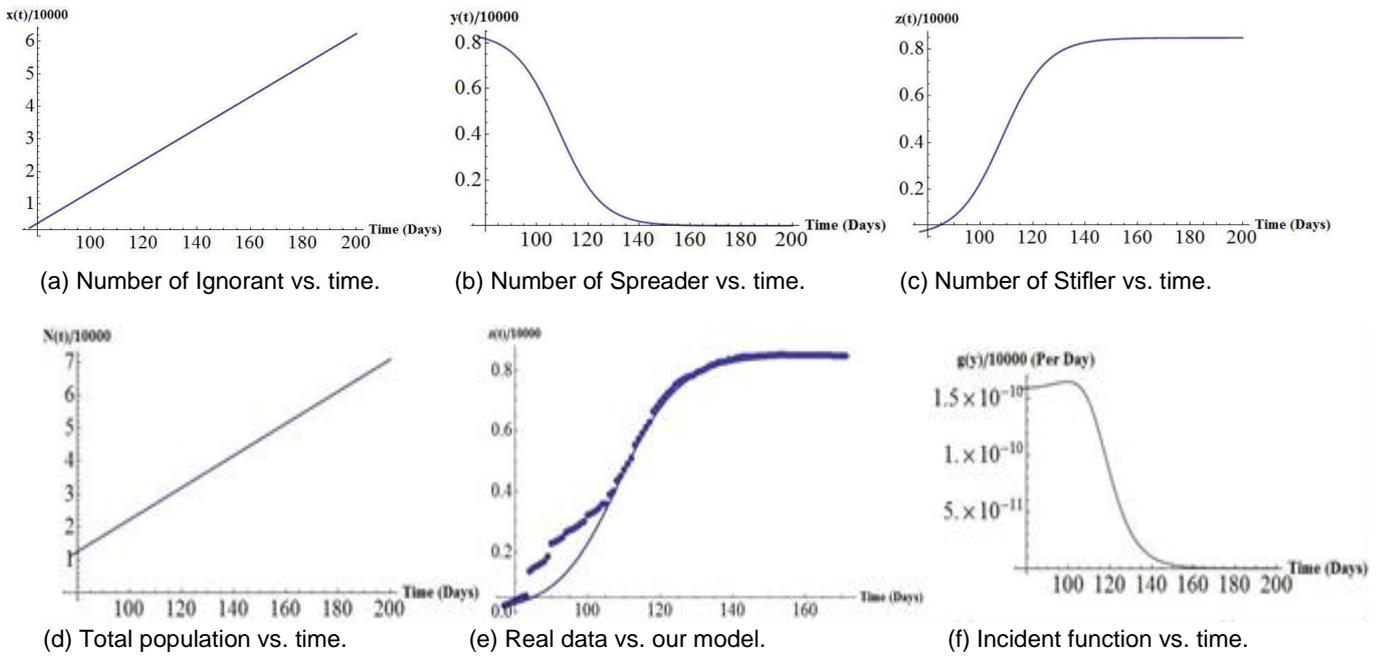

(a) Number of Ignorant vs. time.
(b) Number of Spreader vs. time.
(c) Number of Stifler vs. time.
(d) Total population vs. time.
(e) Real data vs. our model.
(f) Incident function vs. time.

FIG. 11: Results using our model for SARS. The parameter values used in all the plots are Λ = 486 per day, k =5.26 × 10$^{-6}$ per day, α =2.54, μ =1.86 × 10$^{-7}$ per day and λ = 1398 per day, with the initial condition x(77) = 2614, y(77) = 8266 and z(77) = 200. These parameters are generated by fitting 95 real time data points of Ref. 6.